# PERFORMANCE EVOLUTION AND EXPECTATIONS MANAGEMENT: LESSONS FROM TEVATRON AND OTHER MACHINES

V. Shiltsev (Fermilab)

*Abstract*

Below we review the LHC luminosity progress in 2010, discuss the luminosity evolution of the Tevatron collider at different stages of the Collider Runs, emphasize general dynamics of the process, compare with the performance of the other colliders analyze planned and delivered luminosity integrals, and discuss the expectation management lessons.

## INTRODUCTION

In the past 9 months, we witnessed great progress in the LHC commissioning. Luminosity of the 7,000 GeV center of mass energy proton-proton collisions reached $205\times10^{30}$ cm$^{-2}$ s$^{-1}$, or 2% of the design value in less than 6 months – see Fig.1. That level of performance exceeds previous records of the luminosity at CERN set up for 90 GeV electron-positron collisions in LEP at $102\times10^{30}$ cm$^{-2}$ s$^{-1}$ (1998) and for 31 GeV proton-antiproton collisions in ISR at $140\times10^{30}$ cm$^{-2}$ s$^{-1}$ (1982). It is about half of the current Tevatron Run II luminosity record of $402\times10^{30}$ cm$^{-2}$ s$^{-1}$. On average, the LHC luminosity doubled every two weeks, and looking at the plot one could get an impression that if the LHC would not stop proton-proton operation in early November and continue to run and progress "as is", then it could overtake the Tevatron luminosity by mid-November, and will reach its design luminosity sometime mid-2011. So, why nobody believes that could happen?

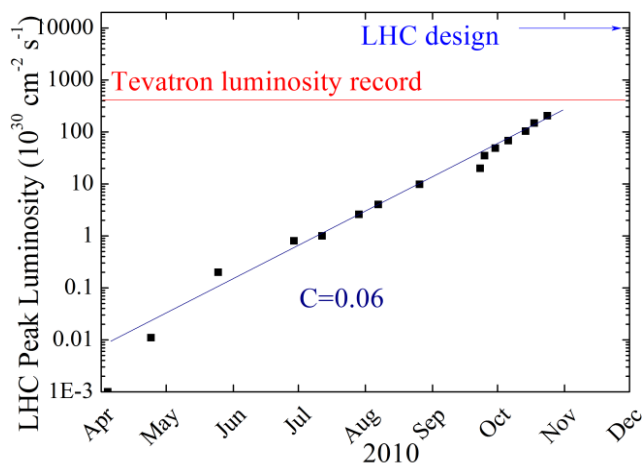

Figure 1: LHC peak luminosity progress in 2010.

The answer is that anyone familiar with performance of other colliders knows that such a fast performance progress is not something unique and unthinkable of at the early stages of the accelerator commissioning. For example, the Tevatron collider had seen luminosity improved by a factor of 12,000 in just three months early in 1987 (that is equivalent doubling the luminosity approximately every week!) and exceeded 1% of then design luminosity of $1.0\times10^{30}$ cm$^{-2}$ s$^{-1}$ [1]. Progress beyond initial stage is determined by careful step-by-step uncovering, analysis and resolution of numerous problems. It was true for all past and present colliders colliders, despite their specific features (species, energies, intensities, etc) and we believe that LHC is not going to be totally unique in that regard – e.g., Table I demonstrates comparable sets of factors of importance for the performance for the Tevatron and LHC.

Table I: Comparison of major factors that play role in the performance progress speed for the Tevatron and LHC.

|  | TEV p-pbar | LHC p-p |
|---|---|---|
| State-of-the-art SC magnets | yes ~800 | yes ~1800 |
| (Old) Sophisticated injector chain | yes 6 | yes 4 |
| Antiproton production/storage/cooling | yes | no |
| Beam-beam effects limiting performance | yes | not yet? |
| Critical importance of collimation | ~no | yes |
| Electron-cloud effects matter | no | yes |
| Space-Charge effects at low energies | yes | yes |

Therefore, it seems reasonable to explore objective laws of the performance evolution of the colliders. In doing that, we will follow the logic of and some data from Reference [2].

## THE TEVATRON PROGRESS AND "CPT THEOREM"

Analysis of the evolution and prediction of high energy colliders' luminosity progress is of great importance for many: it tells machine physicists whether their scientific and technical decisions taken years ago were correct; for the experimental high energy physicists, it is the basis for their schedules and upgrade plans; for the management and funding agencies, it is an important input on the future facilities and projects. The Tevatron luminosity history – see Fig.2 - gives several important lessons in that regard. The luminosity increases occurred after numerous improvements, some of which were implemented during operation, and others introduced during regular shutdown periods.

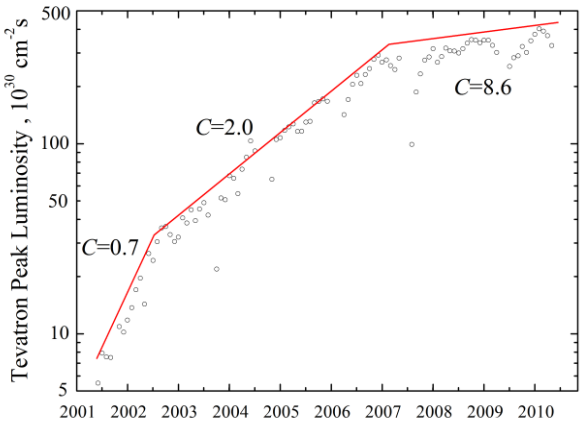

Fig.2: Tevatron peak luminosity progress during Run II (2001-2011).

A large number of improvements (see most of them listed in the Table I in Ref.[2]), took place in all the accelerators of the Collider complex, they were addressing all the parameters affecting the luminosity – proton and antiproton intensities, emittances, optics functions, bunch length, losses, reliability and availability, etc. – and led to fractional increase varying from few % to some 40% with respect to previously achieved level. As the result of some 30 improvements in 2001-2010, the peak luminosity has grown by a factor of about 50 from $L_i \approx 8 \times 10^{30}$ cm$^{-2}$s$^{-1}$ to $L_f \approx 400 \times 10^{30}$ cm$^{-2}$s$^{-1}$, or about 14% per step on average. In principle, such complex percentages - "N% gain per step, step after step, with regular periodicity" - should result in the *exponential* growth of the luminosity $L(t+T)/L(t)=exp(T/C)$. Nevertheless, the pace of the luminosity progress was not always constant. As one can see from Fig.2, the Collider Run II luminosity progress was quite fast with $C \approx 0.7$ year in the period from 2001 to mid-2002 when previous Run I luminosity level was (re)achieved; stayed on a steady exponential increase path with $C \approx 2.0$ yr from 2002 till 2007, and significantly slowed down afterward, $C \approx 8.6$. A plausible hypothesis - so called "CPT theorem for accelerators" - was proposed in Ref.[2], that over extended periods of operation, the performance of colliding facilities evolves in accordance with an approximate formulae:

$$C \cdot P = T \qquad (1)$$

where the factor $P=ln($luminosity$)$ is the "performance" gain over time interval $T$, and $C$ is a machine dependent coefficient equal to average time needed to increase the luminosity by $e=2.71…$ times, or boost the "performance" $P$ by 1 unit. Both, $T$ and $C$ have dimension of time, and the coefficient $C$ was called "complexity" of the machine, as it directly indicates how hard or how easy was/is it to push the performance of individual machine.

In general, the complexity should be dependent on how well understood are physics and technology of the machine.

## EXAMPLES

Let us consider several more illustration of the "CPT theorem for accelerators" – see Figs. 3-10.

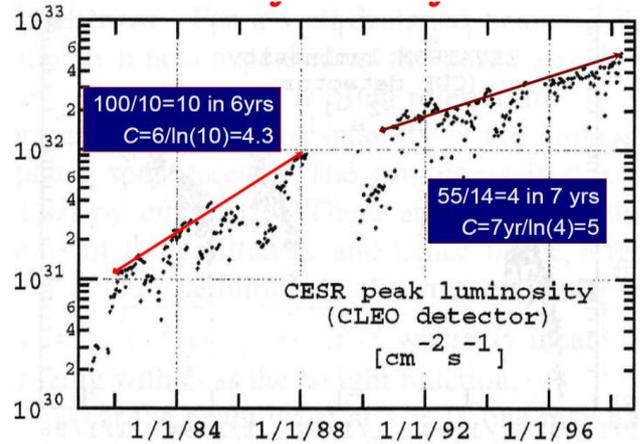

Figure 3: Luminosity history of e+e- collider CESR.

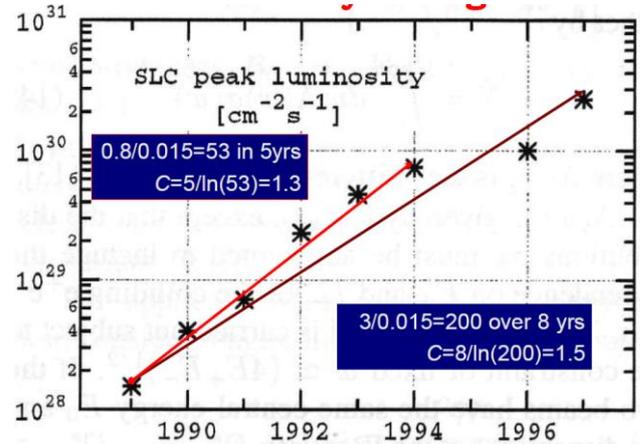

Figure 4: Luminosity history of e+e- linear collider SLC at SLAC.

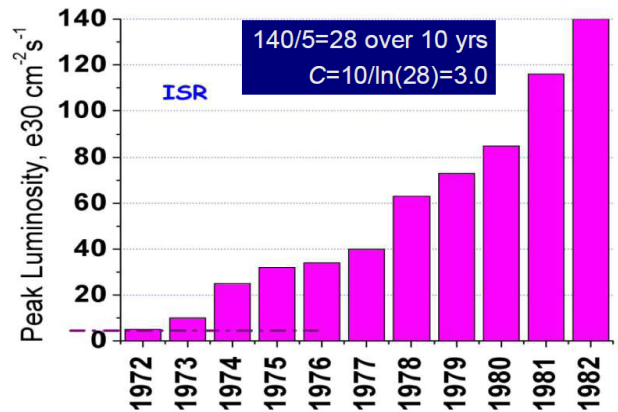

Figure 5: Luminosity history of p-p collider ISR at CERN.

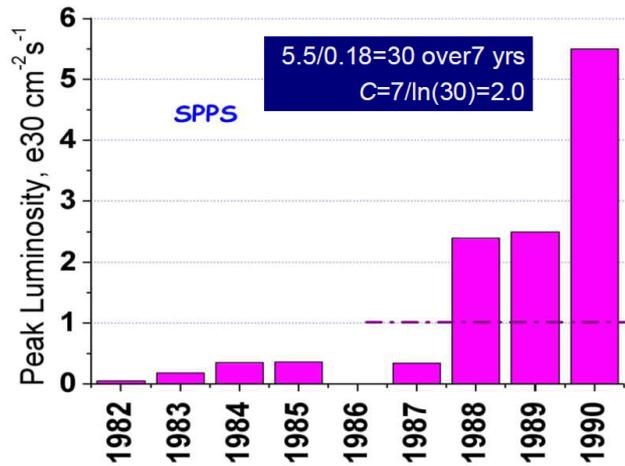

Figure 6: Luminosity history of p-p collider SppS at CERN.

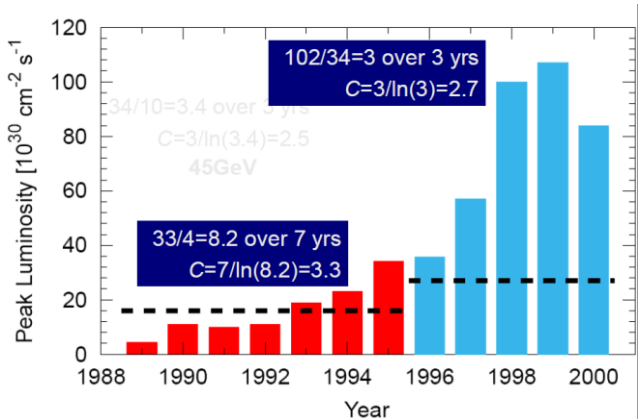

Figure 7: Luminosity history of e+e- collider LEP at CERN.

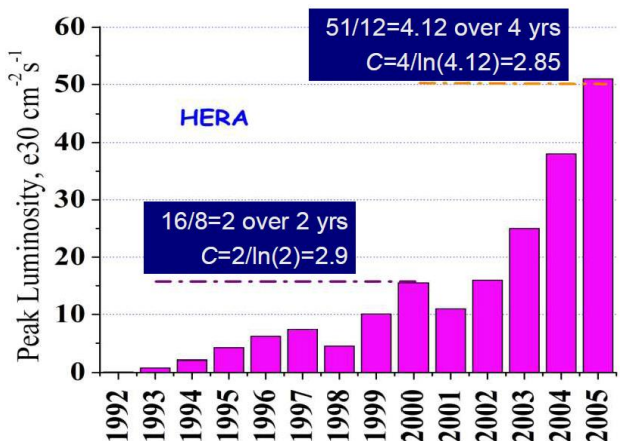

Figure 8: Luminosity history of e-p collider HERA at DESY.

In each figure we indicate – either by straight line or/and by a text box the increase of the record luminosity over certain period of time, and calculate corresponding complexity factors. Table II lists all of them for side by side comparison. One can see that in general, the hadron machines Tevatron, SppS, ISR, HERA and RHIC have average complexity of about $<C>=2.4$ . Effective complexities in the very early periods of operation are very small $C=0.03-0.06$ (Tevatron and LHC).

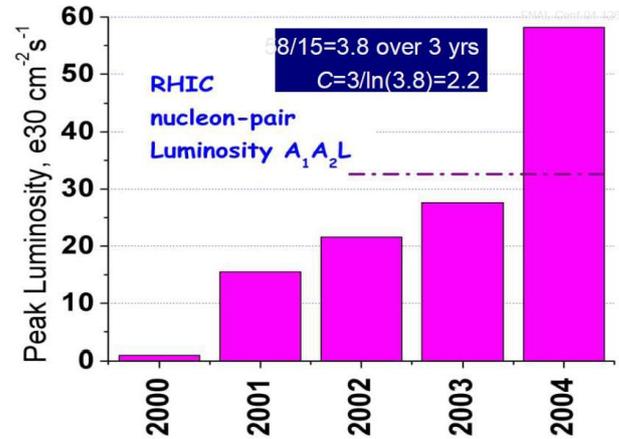

Figure 9: Luminosity history of p-p collider RHIC at BNL.

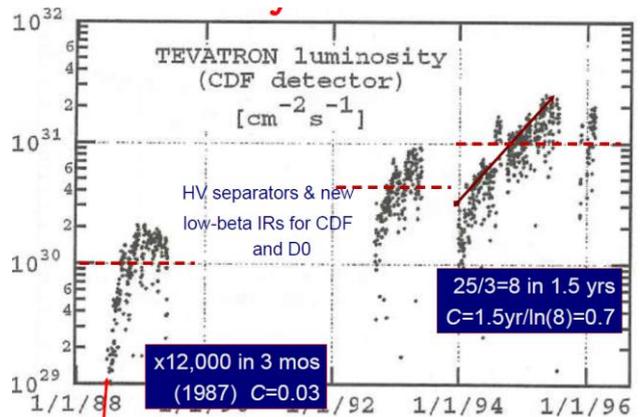

Figure 10: Luminosity history of the Tevatron proton antiproton collider at Fermilab in 1987-1996.

Table II: "Complexities" of colliding beam facilities.

|  | *C* | *years* |
|---|---|---|
| CESR   *e+e-* | **4.3** | 1883-1988 |
| LEP I   *e+e-* | **3.3** | 1989-1995 |
| SLC   *e+e-* | **1.5** | 1989-1997 |
| HERA I, II   *p-e* | **2.9** | 1992-00-2005 |
| ISR   *p-p* | **3.0** | 1972-1982 |
| SppS   *p-pbar* | **2.0** | 1982-1990 |
| Tevatron Run II *p-pbar* | **2.0** | 2002-2007 |
| RHIC   *p-p* | **2.2** | 2000-2004 |
| Tevatron startup | **0.03** | 1987 |
| LHC startup | **0.06** | 2010 |

Differences in machine complexity factors $C$ may be due to various reasons: a) first of all, beam physics issues are quite different not only between classes of machines (hadrons vs e+e-) but often between colliders from the same class – all that affects how fast and what kind of improvements can be implemented; b) accelerator reliability may affect the luminosity progress, especially for larger machines with greater number of potentially not-reliable elements; c) another factor is capability of the team running the machine to cope with challenges, generate ideas for improvements and implement them; d) and, of course, the latter depends on resources available for each team.

Note, that the exponential growth is characteristic to advances in other areas of science and technology. E.g., the maximum energy achieved in particle accelerators grew by factor of 10 every 6 years over many decades [3]. It is often presented in semi-log "Livingston plot" and corresponds to $C=2.6$. Another example is the "Moore's Law" [4] of exponential growth of modern microprocessor speed that doubles every 20 months, yielding $C=2.4$.

## PERFORMANCE EXPECTATION MANAGEMENT

It is well known that the expectations are the only measure of one's success, and that in the case when delivered value is (even slightly) less than expected value – see Fig.11 – then perceived value is much less then delivered one.

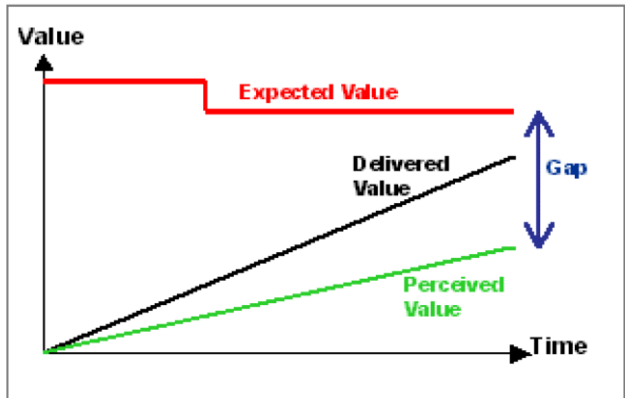

Figure 11: Perceived value of performance.

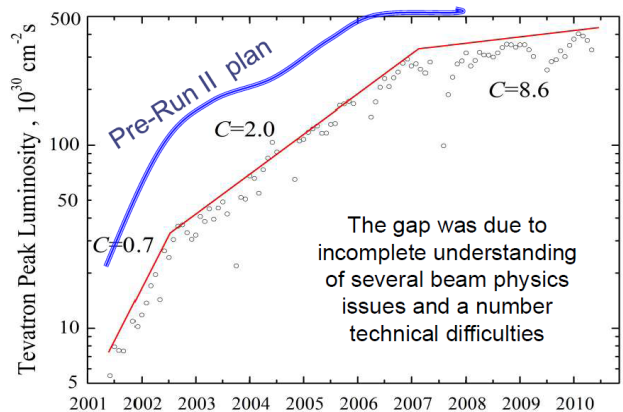

Figure 12: The Tevatron luminosity performance and pre-Run II plan [5].

The Tevatron Run II was a victim of such physiological effect when it was realized that its performance in 2001-2003 was significantly below expectations – see Fig.12. The Run II start-up difficulties were objective (new machine – Main Injector was introduced in operation, Accumulator was greatly upgraded and not optimized, 36 bunches operation was totally new, etc) and a lot of studies were needed to understand and correct them, but, still, as the result the progress seemed unacceptably slow. Only the 2003 DOE review of the Tevatron operation revealed the technical roots of the situation and new approach was embraced: since then, the luminosity goals were expressed in terms of "base" goals that we believe have high degree of certainty of being achieved and "stretched" goals that represent our "best estimate" of the limit of performance to which the facility can be pushed (with the most likely outcome somewhere in between). It is of notion, that careful analysis of the issues and potential progress allowed properly set annual goals, and the Tevatron never missed them since 2003. That had greatly improved predictability of the machine performance and morale of the operating team.

## CONCLUSIONS, RECOMMENDATIONS

One should not expect that the period of incredibly fast growth of the LHC luminosity - as in 2010- will last long. At some point the progress will most probably turn to the rate corresponding to complexity of $C\sim 2$. Such a period of exploration and fight for ultimate performance with $C\approx 2$ might take as short as 3-4 years and as long as 6-10 years It will be followed by relative stabilization of performance (either running out of ideas or preparing for a major upgrade). *A numerical example*: progress from $L=3\times 10^{33}$ cm$^{-2}$s$^{-1}$ to $L=5\times 10^{34}$ cm$^{-2}$s$^{-1}$ might take 6-9 years if $C=2-3$.

Expectations management is crucial. As in the case of the Tevatron, the LHC goals may need to be expressed in terms of two goals: "base" goal – that is believed has very high degree of certainty of being achieved and the "design" or "stretched" goal that represents your "best estimate" of the limit of performance to which the facility can be pushed. The goals and the ratio of "base" to "design" goals will depend on the level of understanding of the machine, e.g. the ratio might change from larger to smaller to reflect lower level of uncertainty in later years.